7

\documentclass[epjc3,twocolumn]{svjour3}          %

\journalname{Eur. Phys. J. C}

\begin{document}

\def\dalam{
\vbox{\hsize 0.5 em \hrule\hbox to 0.5 em{\vrule width 0.5 pt height 0.5 em
\hfill\vrule width 0.5 pt height 0.5 em} \hrule}  }

\title{The investigation of low frequency dilaton generation}

\author{V.I.~Denisov\thanksref{e1,addr1} 
        \and
I.P.~Denisova\thanksref{e2,addr2} 
        \and
      E.T.~Einiev\thanksref{e3,addr2}
}

\thankstext{e1}{e-mail: vid.msu@yandex.ru}
\thankstext{e2}{e-mail: denisovaip@mati.ru}
\thankstext{e3}{e-mail: pm@mati.ru }

\institute{ Department of Physics, Moscow State University, \\ 
119991, Moscow,  Russia
\label{addr1}
\and
Moscow Aviation Institute (National Research University), \\
125993, Moscow, Volokolamskoe Highway 4,  Russia
\label{addr2}
}

\date{Received: date / Accepted: date}

\maketitle

\begin{abstract}
The electromagnetic source of dilaton is a first invariant 
of the electromagnetic field tensor. For 
electromagnetic waves, this invariant can be non zero only in the near zone. 
 Pulsars and magnetars are natural sources of this type.
We calculated the generation of dilatons by coherent 
electromagnetic field of rotating  magnetic dipole  moment
 of pulsars and magnetars. It is shown that the radiation of dilaton waves 
occurs at the two frequencies: the rotation frequency $\omega$
of the magnetic dipole moment of the neutron star and the twice frequency. 

The generation of  dilaton  at frequency $\omega$ 
is maximal in the case when the angle between the 
magnetic dipole moment and the axis of its rotation is $\pi/4.$ 
If this angle is $ \pi/2,$ then dilaton radiation at frequency $\omega$ does not occur.
The generation of  dilaton at frequency $2\omega$ is maximal in the case when 
the angle between the  magnetic dipole and the axis of its rotation is $\pi/2.$ 

Angular distribution of radiation of dilatons having a frequency of $\omega$,
has a maximum along conic surfaces $\theta=\pi/4.$
Angular distribution of radiation of dilatons having a frequency of $2\omega$,
 has a maximum in the plane which is perpendicular to the axis of rotation 
$(\theta=\pi/2).$   
\end{abstract}


\section{Introduction}

In theoretical physics, an interest to the theory of dilatons \cite{1,2} 
has increased again.
In these works, the manifestations of the dilaton field 
in astrophysical and laboratory conditions are theoretically studied.

In particular, the influence of the dilaton scalar field on the properties of strange 
quark stars and quantum deformed Schwarzschild black holes was studied in \cite{3,4}.

In addition, new ideas about the diversification of the theory
in the field of modern physical experiment have appeared.
It has led to a necessity of carrying on an extra 
analytical research of dilatons behavior  in a significantly non-liner sector 
\cite{5,6,7,8} and using  fully  nonlinear, numerical investigation \cite{9,10}
as well as at high energies  available for LHC experiments.  

Particularly, in paper \cite{11} authors have performed a detailed study 
of dilatons  phenomenology 
 in a composite twin Higgs model; the authors of article \cite{12} explore the 
possibility that a light dilaton can be the first sign of new physics at 
the LHC.

However,  there are such processes of dilatons generation  that are available and
at low energies. This paper is devoted to one of these processes.

The density of the Lagrange function in Maxwell-dilaton theory, following 
the work \cite{9}, 
we write in the form:
$${\cal L} =a_0(\partial\Psi)^2+a_1e^{-2{\cal K}\Psi}F^{nm}F_{nm},\eqno(1)$$
where $\Psi$ is the scalar field of the dilaton, $F_{nm}$ is the 
electromagnetic 
field tensor, $a_0,\ a_1$ and $\cal K$ is the coupling constants.


It should be noted that recently the concept of dilaton has been expanded. 
In effective field theory \cite{11,12,13,14}, a dilaton is a scalar field, whose 
Lagrangian density, in the particular 
case of interaction with an electromagnetic field, coincides with the Lagrangian density (1), 
and has no relation to the multidimensional theory of gravity. 
Therefore, in this paper we will use the Lagrangian density (1) without discussing 
the physical nature of the dilaton.

In pseudo-Euclidean space-time, the field equations
obtained from the density of the Lagrange function (1) have the form:
$$\dalam \Psi=\frac{a_1{\cal K}}{ a_0}e^{-2{\cal K}\Psi} F_{nm}F^{nm}
=\frac{2a_1{\cal K}}{ a_0} 
e^{-2{\cal K}\Psi}\big[B^2-E^2\big],\eqno(2)$$
where $\bf B$ and $\bf E$ are the induction  of the magnetic field 
and, the intensity of the electric field, that create the dilaton field.

The dynamics of the Maxwell-dilaton theory (2) in Minkowski spacetime  
we will only consider in the weak dilaton field approximation: 
$|{\cal K}\Psi|<<1.$ 
In this case, the equation for the dilaton field will take the form:
$$\dalam \Psi=\frac{2a_1{\cal K}}{ a_0} \big[B^2-E^2\big].\eqno(3)$$

Thus, for $|{\cal K}\Psi|<<1$, the source of dilaton radiation is
only the first invariant of the electromagnetic field tensor $F_{ik}.$
Since this invariant is zero in the wave zone of any electromagnetic waves,
the noticeable radiation of dilatons is possible only from the near zone of 
electromagnetic 
waves,where this invariant is not zero.
Therefore, the effective electromagnetic generators
of dilatons are coherent electromagnetic waves, in which
in the non-wave zone, the fields $\bf E$ and $\bf B$ satisfy the condition:
${\bf B}^2\neq {\bf E}^2$. 
These are the properties of the magnetic dipole radiation of pulsars and 
magnetars.
They have strong magnetic fields, that comparable and even  exceed 
the quantum field $B_q=4.41\cdot 10^{13}$ gauss: 
the induction of a magnetic dipole field on the surface of a pulsar can reach
$10^{13}$  gauss \cite{15}, and on the surface of a magnetar -- up to 
$2\cdot10^{15}$ gauss \cite{16}.

Therefore, the first invariant of the electromagnetic field $F_{nm}F^{nm}$
for the magnetic dipole radiation of pulsars and magnetars in the near zone 
takes 
on the value, that can hardly be created in other electromagnetic processes.

\section{Calculation of dilaton generation by magnetic dipole
radiation of pulsars and magnetars}

       Consider a pulsar or magnetar of radius $R_s$ with a magnetic dipole 
moment $\bf M$ rotating 
with a frequency of $\omega$ around an axis making an angle $\alpha$ with 
a vector $\bf M$.

Then the pulsar magnetic dipole moment has the components:
$${\bf M}(\tau)=|{\bf M}|(\cos(\omega \tau)\sin\alpha, \ 
\sin(\omega \tau)\sin\alpha,\ \cos\alpha),$$
where $\tau$ is retarded time: $\tau=t-r/c$.

Due to the rotation of the vector $\bf M$, magnetic dipole radiation of 
electromagnetic waves 
is being generated. According to work \cite{17}, 
this radiation vectors $\bf B$ and  $\bf E$,  can be written 
in the form: 
$${\bf B}({\bf r},\tau)
=\frac{3({\bf M}(\tau)\cdot {\bf r}){\bf r}-r^2{\bf M}(\tau)}{ r^5}
-\frac{{ \dot{\bf M}}(\tau)}{ c r^2}+
\eqno(4)$$
$$+\frac{3({ \dot{\bf M}}(\tau)\cdot {\bf r}){\bf r}}{ c r^4}
+\frac{(\ddot{\bf M}(\tau)\cdot {\bf r}){\bf r}-r^2\ddot{\bf M}(\tau)}{ c^2 r^3},$$
$${\bf E}({\bf r},\tau)=\frac{({\bf r}\times\dot{\bf M}(\tau))}{ c r^3}+
\frac{({\bf r}\times\ddot{\bf M}(\tau))}{ c^2 r^2},$$
where the dot above the vector ${\bf M}$ means the derivative on retarded time $\tau$. 

For pulsars and magnetars condition $\omega R_s<<c,$ are met, hence they are 
in the near 
zone of their own magnetic dipole radiation.

The total intensity 
$I$ -- the amount of energy of the 
electromagnetic waves emitted in all directions by a rotating magnetic 
dipole per time unit:
$$I_{EMW}=\frac{2\ddot {\bf M}^2}{ 3c^3}=\frac{2cB_0^2R_s^6k^4}{ 3}\sin^2\alpha.
\eqno(5)$$
where  $k=\omega/c$ and the square of the magnetic dipole moment vector of the 
neutron star $ M^2 $ 
is written in terms of the square of the magnetic induction vector on the 
surface 
of the star $ B_0^2 $ according to the equation: $ M^2=B_0^2R_s^6. $
          
       Using relations (4), and keeping only the time-dependent part,
we calculate the invariant ${\bf B}^2-{\bf E}^2:$ 

$${\bf B}^2-{\bf E}^2=\frac{B_0^2R_s^6}{ 2 r^6}\Big\{\sin^2\alpha\sin^2\theta
\Big[(3-2k^2r^2)\times\eqno(6)$$
$$\times\cos\big[2(\varphi+kr-\omega t)\big]+6kr\sin\big[2(\varphi+kr-\omega t)\big]\Big]+$$
$$+2\sin2\alpha\sin\theta\cos\theta\Big[
(3+k^2r^2)
\cos\big[2(\varphi+kr-\omega t)\big]+$$
$$+3kr\sin\big[2(\varphi+kr-\omega t)\big]
\Big]\Big\}.
$$

Substituting expressions (6) in the right part of equation (3), for the 
convenience of 
the solution, we will rewrite it in a complex form, assuming that after 
solving it, we will 
leave only the real part. Then we have
$$\Delta\ \Psi(\vec r,t)-\frac{1}{ c^2}\frac{\partial^2 \Psi(\vec r,t)
}{ \partial t^2}
=\frac{a_1{\cal K}  B_0^2R_s^6}{ a_0r^6}\Big\{\sin^2\alpha 
\sin^2\theta\Big[3-$$
$$-2k^2r^2+6ikr\Big]e^{-2i(\varphi+kr-\omega t)}-
\eqno(7)$$
$$-2\sin2\alpha \sin\theta\cos\theta  \Big[(3+k^2r^2)
+3ikr\Big]e^{-i(\varphi+kr-\omega t)}\Big\}. $$

The retarded solution \cite{18} of the equation (7) has the form:
$$\Psi(\vec r,t)=-\frac{ a_1 B_0^2R_s^6}{ 8\pi a_0}\int\limits_V
\frac{dV'}{ r'^6|r-r'|}
\Big\{\sin^2\alpha \sin^2\theta'\Big[3-
\eqno(8)$$
$$-2k^2r'^2+6ikr'\Big]e^{-2i(\varphi'+kr'-\omega t+k|r-r'|)}+$$
$$+2\sin2\alpha \sin\theta'\cos\theta'\Big[(3+k^2r'^2)+3ikr'\Big]\times$$
$$\times e^{-i(\varphi'+kr'-\omega t+k|r-r'|)}\Big\},$$
where 
${\bf r'}=(r'\sin\theta'\cos\varphi', r'\sin\theta'\sin\varphi',r'\cos\theta')$
and   
${\bf r}=(r\sin\theta\cos\varphi, r\sin\theta\sin\varphi,r\cos\theta).$

Using the Gegenbauer theorem \cite{19} and the formulas of the article \cite{20}, we 
find the field
$\Psi(r,t)$ in the region $r>R_s$ (see Appendix A):
$$\Psi(\vec r,t)=\frac{i\pi^2  B_0^2R_s^6}{ \sqrt{r}}e^{2i\omega t}\sin^2\alpha
            \sin^2\theta  e^{-2i\varphi}\times\eqno(9)$$ 
$$\times\Big\{H^{(2)}_{5/2}(2kr)[g_1(2kr)-g_1(2kR_s)]
-J_{5/2}(2kr)g_2(2kr)\Big\}+$$
$$+\frac{i\pi a_1{\cal K}  B_0^2R_s^6}{ a_0\sqrt{r}}e^{i\omega t}\sin2\alpha
\sin\theta \cos\theta e^{-i\varphi}
\times$$
$$\times\Big\{H^{(2)}_{5/2}(kr)[g_3(kr)-g_3(kR_s)]-J_{5/2}(kr)g_4(kr)\Big\},$$
where the notations are entered:
$$g_1(z)=\sqrt{\frac{k^9}{ 2\pi}}\Big\{\frac{(2z^3+3iz^2+12z-6i)\exp(-2iz)}{ 4z^6}+$$
$$+\frac{8z^3-2iz^4+9iz^2+6i}{ 4z^6}\Big\},$$
$$g_2(z)=\sqrt{\frac{k^9}{ 2\pi}}\frac{[2z^3+3iz^2+12z-6i]\exp(-2iz)
}{ 2z^6},$$
$$g_3(z)=\frac{2\sqrt{k^9}[(15iz^2-2z^3+24z-12i)\exp(-2iz)]}{ \sqrt{\pi}z^6}+$$
$$+\frac{2\sqrt{k^9}[2iz^4+4z^3+9iz^2+12i]}{ \sqrt{\pi}z^6},$$
$$g_4(z)=-\frac{4k^{9/2}[2z^3-15iz^2-24z+12i]\exp(-2iz)
}{ \sqrt{\pi}z^6}.$$
The expression (9) is an exact solution of the equation (7).

\section{The angular distribution  of the dilaton radiation}

Let us  study the angular distribution
 of the arising dilaton radiation. By 
definition \cite{18}, the amount of energy $dI$, emitted by the source per unit 
time through the 
solid angle $d\Omega=\sin\theta d\theta d\varphi,$ is given by the formula:
$$\frac{dI}{ d\Omega} =\lim_{r\to\infty}r({\bf W\cdot r}),$$              
where vector $\bf W$ in tensor form has  the components  
$W^\alpha=cT^{0\alpha},$
and $T^{nm}$ is the energy momentum tensor of dilaton radiation.

The energy momentum tensor of the dilaton field has the form:
$$T^{ik}=2a_0g^{in}g^{km}\big\{\frac{\partial \Psi}{ \partial x^n}
\frac{\partial \Psi}{ \partial x^m}
-\frac{1}{ 2}g_{nm}\frac{\partial \Psi}{ \partial x^k}
\frac{\partial \Psi}{ \partial x^p} g^{kp}\big\}.$$
Then the angular distribution  of the dilaton radiation
 will be determined by the expression: 
$$\frac{dI}{ d\Omega}=-2a_0\lim\limits_{r\to \infty}r
(\vec r\ \vec \nabla \Psi)\frac{\partial \Psi}{\partial t}.\eqno(10)$$
For further investigation of the generation of dilaton radiation by the 
electromagnetic field of a rotating magnetic dipole, we only need the wave 
part of  
expression (9), which decreases at $kr>>1$ as  $1/r.$ 
In addition, we take into account that for
most pulsars and magnetars $kR_s<<1,$ and keeping in the resulting expression
only the asymptotically main term in the expansions with respect to this small 
parameter.
Then, keeping in expression (9) the real part and discarding the non-wave 
terms, we get
 (see Appendix B):
$$\Psi(\vec r,t)=\frac{2a_1{\cal K}  B_0^2k^2R_s^5}{ 5a_0r}
\Big\{2\sin^2\alpha\sin^2\theta\cos[2(\omega t-kr-\varphi)]-\eqno(11)$$
$$-\sin2\alpha\sin\theta \cos\theta \cos(\omega t-kr-\varphi)\Big\}.$$

Substituting the expression (11) into the ratio (10) and averaging the 
resulting formula
for the wave period $T=2\pi/\omega$, we come to the expression:

 $$\frac{dI}{ d\Omega}=\frac{dI}{ d\Omega}(\omega)
+\frac{dI}{ d\Omega}(2\omega),$$
where
$$\frac{dI}{ d\Omega}(\omega)=\frac{4ca_1^2{\cal K}^2B_0^4k^6R_s^{10}}{ 25a_0}
\sin^22\alpha\sin^2\theta\cos^2\theta\eqno(12)
$$ is the angular distribution of the radiation of dilatons having the 
frequency $\omega$, and
$$\frac{dI}{ d\Omega}(2\omega)=\frac{64ca_1^2{\cal K}^2B_0^4k^6R_s^{10}}{ 25a_0}
\sin^4\alpha\sin^4\theta\eqno(13)
$$is the angular distribution of the radiation of dilatons having the frequency $2\omega$.

Integrating  these expressions over the angles $\theta $ 
and $\varphi$, we obtain the 
total intensity  $I$ -- the amount of energy of the 
dilatonic waves emitted in all directions by a rotating magnetic dipole per 
time unit: 
$$I(\omega)=\frac{32\pi ca_1^2{\cal K}^2B_0^4k^6R_s^{10}}{ 375a_0}
\sin^22\alpha,\eqno(14)$$
$$I(2\omega)=\frac{2048\pi ca_1^2{\cal K}^2B_0^4k^6R_s^{10}}{ 375a_0}
\sin^4\alpha.$$
It follows from the expressions (12)-(14) that both the angular distributions 
and the total 
intensities of dilaton radiation at the frequencies $2 \omega$ and $\omega$ 
differ 
significantly, although they are generated by the same electromagnetic fields 
(4) of 
the rotating magnetic dipole of a neutron star.

\section{Discussion}

  The calculation showed that dilaton radiation  in general occurs at the two 
frequencies:
the rotation frequency $\omega$
of the magnetic dipole moment of the neutron star and the twice frequency 
$2\omega$. 

Angular distribution of radiation of dilatons having a frequency of $\omega$,
has a maximum along conic surfaces $\theta=\pi/4.$

Angular distribution of radiation of dilatons having a frequency of $2\omega$,
 has a maximum in the plane which is perpendicular to the axis of rotation 
$(\theta=\pi/2).$   

As it follows from the expression (12),  the generation of  dilaton  at 
frequency $\omega$ 
is maximal in the case when the angle between the 
magnetic dipole moment and the axis of its rotation is $\pi/4.$ 
If this angle is $ \pi/2,$ then dilaton radiation at frequency $\omega$ does 
not occur.

The generation of  dilaton at frequency $2\omega$ is maximal in the case when 
the angle between the  magnetic dipole moment and the axis of its rotation 
is $\pi/2.$ 

Therefore, the newly discovered "Magnificent Seven" magnetars  \cite{21,22}
should emit dilatons only at the frequency 
$2 \omega$, since dilatonic 
radiation at  frequency $\omega$ 
 is either absent or strongly suppressed, since they have the angle between the 
magnetic dipole moment and the rotation axis close to $\pi/2.$

It should be noted that the dipole radiation generated by the rotation of the magnetic 
dipole moment of pulsars and magnetars is also a source of generation of two types of  
axion-like particles: massive axions and strictly massless arions.
Unlike dilatons, the electromagnetic source for which is the invariant 
$F_{nk}F^{nk}=2(B^2-E^2)$ of the 
electromagnetic field tensor, the source of axion-like particles is the pseudo-invariant  
$({\bf B\ E})$.

Therefore the radiation of arions, as shown in \cite{23}, occurs only at the rotation 
frequency $\omega$. 

Currently, the values of the constants $ a_0,\ a_1$ and $\cal K$ are unknown.

Let's roughly estimate the values of combination $\eta=a_1^2{\cal K}^2/a_0$
these constants.  
To do this, it is reasonable to require that the maximum value of the intensity
of the dilaton radiation (14) was significantly less
than the maximum value of the intensity
of electromagnetic radiation (5):
$$\hbox{max}\ I_{\hbox{Dilaton}}<<\hbox{max}\ I_{\hbox{EMW}}.$$
We suppouse, that the dilaton radiation, like any physical radiation
carries non-negative energy, then the constant $a_0>0.$
From expressions (5) and (14) in order of magnitude, we get the inequality:
$$\eta<<k^{-2}R_s^{-4}B_s^{-2}.\eqno(15)$$
Let's see how the right-hand side of this inequality changes for pulsars and
magnetars.

Consider the pulsar PSR J1810+1744. According to \cite{15}, its radius is 
estimated as  $\sim 10^6$ cm, the period of its rotation is 1.6 milliseconds, 
and the magnetic field induction on the surface is about $10^{13}$ gauss.
In this case, the expression (15) takes the form:
$$\eta<<10^{-36}\ \hbox{cm}\hbox{ erg}^{-1}.$$

Consider now magnetar   MG J1647-4552.  According to \cite{16}, its radius is 
estimated as  $\sim 10^6$ cm, the period of its rotation is 10 seconds, and 
the magnetic field 
induction on the surface is about $10^{15}$  gauss.
In this case, the expression (15) takes the form:
$$\eta<<10^{-32}\ \hbox{cm}\hbox{ erg}^{-1}.\eqno(16)$$
Thus, pulsars give a tighter estimate by the value of $ \eta$ than magnetars, 
as they are more intense sources of dilaton radiation.

Note that in the system of units of measurement, that we use here, the unit 
for the
density of the Lagrange function is $\hbox{erg}\ \hbox{cm}^{-3},$ 
the dilaton field $\Psi$ and constants $a_1$ and ${\cal K}$ are dimentionless,
the unit for the constant $a_0$ is $\hbox{erg}\ \hbox{cm}^{-1}$, the unit for 
the magnetic field 
$B$ is $\hbox{erg}^{1/2}\hbox{ cm}^{-3/2}.$                          

If we use a system of units in which $c=\hbar=1$, then the inequality (16) 
takes the form:
$$\eta<<10^{-47}\ \hbox{Gev}^{-2}.$$
 
\vskip 6 true mm

{\bf{Appendix A. Calculation of dilaton generation}}

\vskip 6 true mm

According to Gegenbauer's theorem \cite{19}, the function 
$\exp\{- ik|{\bf r}-{\bf r'}|\}/|{\bf r}-{\bf r'}|$
can be decomposed into an infinite series by Legendre polynomials from 
the cosine of 
the angle between the vectors $\bf r'$ and ${\bf r}.$

For $r>r'$, this series has the form:
$$\frac{\exp\{-ik|{\bf r}-{\bf r'}|\}}{ |{\bf r}-{\bf r'}|}=\eqno(A1)$$
$$=-\frac{\pi i}{ 2\sqrt{rr'}}\sum\limits_{n=0}^\infty (2n+1)J_{n+1/2}(kr')
H^{(2)}_{n+1/2}(kr)P_n(\cos\gamma),$$
and when $r<r'$
$$\frac{\exp\{-ik|{\bf r}-{\bf r'}|\}}{ |{\bf r}-{\bf r'}|}=\eqno(A2)$$
$$=-\frac{\pi i}{ 2\sqrt{rr'}}\sum\limits_{n=0}^\infty (2n+1)J_{n+1/2}(kr)
H^{(2)}_{n+1/2}(kr')P_n(\cos\gamma),$$
where 
$$\cos\gamma=\cos\theta\cos\theta'
+\sin\theta\sin\theta'\cos(\varphi'-\varphi).$$

Substituting the expressions (A1-A2) in the retarded integral (8), we bring 
it to the form:
$$\Psi(\vec r,t)=\frac{ia_1{\cal K} B_0^2R_s^6}{ 8a_0\sqrt{r}}e^{2i\omega t}
\sin^2\alpha
\sum\limits_{n=0}^\infty (2n+1)
\int\limits_0^\pi \sin^3\theta'd\theta'\times \eqno(A3)  $$
$$\times\int\limits_0^{2\pi} e^{-2i\varphi'}P_n(\cos\gamma)d\varphi'
\Big\{H^{(2)}_{n+1/2}(2kr)\int\limits_{R_s}^r 
\frac{1}{ r'^{9/2}}\Big[3-$$
$$-2k^2r'^2+6ikr'\Big]e^{-2ikr'}J_{n+1/2}(2kr')dr'+$$
$$+J_{n+1/2}(2kr)\int\limits_r^\infty 
\frac{1}{ r'^{9/2}}\Big[3-$$
$$-2k^2r'^2+6ikr'\Big]e^{-2ikr'}H^{(2)}_{n+1/2}(2kr')dr'
\Big\}+$$
$$+\frac{i\pi B_0^2R_s^6}{ 2\sqrt{r}}e^{i\omega t}\sin2\alpha
\sum\limits_{n=0}^\infty (2n+1)
\int\limits_0^\pi \sin^2\theta' \cos\theta' 
\int\limits_0^{2\pi} e^{-i\varphi'}
\times$$
$$\times P_n(\cos\gamma) d\theta'd\varphi'\Big\{H^{(2)}_{n+1/2}(kr)\int\limits_{R_s}^r 
\frac{1}{ r'^{9/2}}\Big[3+$$
$$+k^2r'^2+3ikr'\Big]e^{-ikr'}J_{n+1/2}(kr')dr'+$$
$$+J_{n+1/2}(kr)\int\limits_r^\infty 
\frac{1}{ r'^{9/2}}\Big[3+$$
$$+k^2r'^2+3ikr'\Big]e^{-ikr'}H^{(2)}_{n+1/2}(kr')dr'
\Big\}.$$

Using the notation
$$N^1=\sin\theta\cos\varphi,\ N^2=\sin\theta\sin\varphi,\ N^3=\cos\theta $$ 
of the article \cite{20}, we can write
$$\sin^2\theta' e^{-2i\varphi'}=(N'^1)^2-(N'^2)^2-2iN'^1N'^2,$$
$$\sin\theta' \cos\theta' e^{-i\varphi'}=(N'^1-iN'^2)N'^3.$$
Substituting these expressions in equation (A3), and given that according 
to article \cite{20} 
$$\int\limits_0^\pi \sin\theta d\theta'\int\limits_0^{2\pi}N^\alpha N^\beta 
P_n(\cos\gamma)
d\varphi'=\frac{4\pi}{ 3}\Big\{\delta^{\alpha\beta}\delta_{0n}+$$
$$+\frac{1}{ 5}\delta_{2 n}\Big[3 N^\alpha N^\beta-\delta^{\alpha\beta}\Big]
\Big\},$$
we get after integrating (A3) over the angles $\theta'$ and $\varphi':$
$$\Psi(\vec r,t)=\frac{i\pi a_1{\cal K} B_0^2R_s^6}{ 2a_0\sqrt{r}}e^{2i\omega t}
\sin^2\alpha
            \sin^2\theta  e^{-2i\varphi}\times\eqno(A4)$$
$$\times\Big\{H^{(2)}_{5/2}(2kr)\int\limits_{R_s}^r 
\frac{1}{ r'^{9/2}}\Big[3-2k^2r'^2+6ikr'\Big]e^{-2ikr'}\times $$
$$\times J_{5/2}(2kr')dr'+J_{5/2}(2kr)\int\limits_r^\infty 
\frac{1}{ r'^{9/2}}\Big[3-2k^2r'^2+$$
$$+6ikr'\Big]e^{-2ikr'}H^{(2)}_{5/2}(2kr')dr'\Big\}+$$
$$+\frac{i\pi a_1{\cal K} B_0^2R_s^6}{ a_0\sqrt{r}}e^{i\omega t}\sin2\alpha
\sin\theta \cos\theta e^{-i\varphi}
\times$$
$$\times\Big\{H^{(2)}_{5/2}(kr)\int\limits_{R_s}^r 
\frac{J_{5/2}(kr')}{ r'^{9/2}}\Big[3+k^2r'^2
+3ikr'\Big]e^{-ikr'}dr'+$$
$$+J_{5/2}(kr)\int\limits_r^\infty 
\frac{H^{(2)}_{5/2}(kr')}{ r'^{9/2}}\Big[3+k^2r'^2+3ikr'\Big]e^{-ikr'}dr'
\Big\}.$$

Let note that
$$J_{5/2}(w)=\sqrt{\frac{2}{ \pi w}}\Big\{\Big[\frac{3}{ w^2}-1\Big]\sin w
-\frac{3\cos w}{ w}\Big\}, $$
$$H^{(2)}_{5/2}(w)=\sqrt{\frac{2}{ \pi w}}\Big\{i\Big[\frac{3}{ w^2}-1\Big]
-\frac{3}{ w}\Big\}e^{-iw}. 
$$
Substituting these relations for the integrals in the expression (A4) 
and integrating them 
in parts, we have:
$$\Psi(\vec r,t)=\frac{i\pi^2 B_0^2R_s^6}{ \sqrt{r}}e^{2i\omega t}\sin^2\alpha
            \sin^2\theta  e^{-2i\varphi}\times \eqno(A5)$$ 
$$\times\Big\{H^{(2)}_{5/2}(2kr)[g_1(2kr)-g_1(2kR_s)]
-J_{5/2}(2kr)g_2(2kr)\Big\}+$$
$$+\frac{i\pi a_1{\cal K} B_0^2R_s^6}{ a_0\sqrt{r}}e^{i\omega t}\sin2\alpha
\sin\theta \cos\theta e^{-i\varphi}
\times$$
$$\times\Big\{H^{(2)}_{5/2}(kr)[g_3(kr)-g_3(kR_s)]-J_{5/2}(kr)g_4(kr)\Big\},$$
where the notation is used:
$$g_1(z)=\sqrt{\frac{k^9}{ 2\pi}}\Big\{\frac{(2z^3+3iz^2+12z-6i)\exp(-2iz)}{ 4z^6}+$$
$$+\frac{8z^3-2iz^4+9iz^2+6i}{ 4z^6}\Big\},$$
$$g_2(z)=\sqrt{\frac{k^9}{ 2\pi}}\frac{[2z^3+3iz^2+12z-6i]\exp(-2iz)
}{ 2z^6},$$
$$g_3(z)=\frac{2\sqrt{k^9}[(15iz^2-2z^3+24z-12i)\exp(-2iz)]}{ \sqrt{\pi}z^6}+$$
$$+\frac{2\sqrt{k^9}[2iz^4+4z^3+9iz^2+12i]}{ \sqrt{\pi}z^6},$$
$$g_4(z)=-\frac{4k^{9/2}[2z^3-15iz^2-24z+12i]\exp(-2iz)
}{ \sqrt{\pi}z^6},$$

\eject

\vskip 6 true mm

{\bf Appendix B. Dilaton field in the wave zone}
\vskip 6 true mm
Let us now construct the asymptotically main part of the dilaton field in 
the wave zone, 
i.e., at $kr>>1$.
In this zone, we have the asymptotics:
$$\frac{H^{(2)}_{5/2}(2kr)}{ \sqrt{r}}=-\frac{i}{ r\sqrt{\pi k}}e^{-2ikr},\eqno(B1)$$ 
$$H^{(2)}_{5/2}(x)=-i\sqrt{\frac{2}{ \pi x}}e^{-ix}, \ \ 
\frac{H^{(2)}_{5/2}(kr)}{ \sqrt{r}}=-\frac{i\sqrt{2}}{ r\sqrt{\pi k}}e^{-ikr}.$$
$$g_1(2kr)\sim\frac{1}{ r^2},\ g_2(2kr)\sim\frac{1}{ r^3},\
g_3(kr)\sim\frac{1}{ r^2},\ g_4(kr)\sim\frac{1}{ r^3}.$$

Since $kR_s<<1$, the function $g_1(2 kR_s)$ and $g_3(kR_s)$
we leave only the asymptotically principal parts.

From the relations (A5), it follows that for $z<<1$, the estimates are valid:
$$ \lim\limits_{z\to 0}zg_1(z)=-\frac{16k^{7/2}}{ 5\sqrt{\pi}},\ \ 
\lim\limits_{z\to 0}zg_3(z)=-\frac{2k^{7/2}}{ 5\sqrt{2\pi}}.\eqno(B2)$$
Therefore
$$g_1(2kR_s)=-\frac{8k^{5/2}}{ 5\sqrt{\pi}R_s},\ \
g_3(kR_s)=-\frac{2k ^{5/2}}{ 5\sqrt{2\pi}R_s}. $$

The expression (9), taking into account the relations (B1)-(B2), takes 
the form:
$$\Psi(\vec r,t)=
\frac{2a_1{\cal K} B_0^2k^2R_s^5}{ 5a_0r}
\Big\{2\sin^2\alpha\sin^2\theta e^{2i(\omega t-kr-\varphi)}-$$
$$-\sin2\alpha\sin\theta \cos\theta e^{i(\omega t-kr-\varphi)}\Big\}.$$
The real part of this expression has the form:
$$\Psi(\vec r,t)=
\frac{2a_1{\cal K} B_0^2k^2R_s^5}{ 5a_0r}
\Big\{2\sin^2\alpha\sin^2\theta\cos[2(\omega t-kr-\varphi)]-$$
$$-\sin2\alpha\sin\theta \cos\theta \cos(\omega t-kr-\varphi)\Big\}.$$
This expression is dilaton field in the wave zone.

\end{document}